\documentclass[dvips]{article}
\usepackage[left]{lineno}
\usepackage{icrctc07}

\title{Search for large-scale anisotropies with the Auger Observatory}
\shorttitle{Search for large-scale anisotropies with the Auger Observatory}

\authors{E. Armengaud, for the Pierre Auger Collaboration}
\shortauthors{E. Armengaud, for the Pierre Auger Collaboration}
\afiliations{Pierre Auger Observatory, Av. San Martin Norte 304, (5613) Malarg\"ue, Argentina}
\email{eric.armengaud@cea.fr}

\abstract{We use more than two years of data from the Pierre Auger Observatory to search for anisotropies on large scales in different energy windows. We account for various systematics in the acceptance, in particular due to the array growth and weather variations. We present the results of analyses and consistency checks looking for patterns in the right ascension modulation of the cosmic ray distribution. No significant anisotropies of this kind are observed.}

\begin{document}
\maketitle

\section{Introduction}
The Pierre Auger Observatory has accumulated an important statistics of cosmic rays (CRs) above $10^{18}\,\mbox{eV} = 1\,$EeV. The use of water-Cherenkov tanks allows to observe a large fraction of the sky. Upper limits were set on the possible fluxes from localized regions near the galactic center~\cite{gc-paper}. 
It is not known yet whether the sources of cosmic rays with an energy just below the ankle are galactic, or whether they have a cosmological distribution~\cite{ankle-papers-a,ankle-papers-b}. If the origin of cosmic rays below the ankle region is galactic, their diffusive escape from the galaxy may be efficient enough so that the sky distribution of $10^{18}$ eV cosmic rays is not completely isotropised as seen from the Earth. The predictions for the shape and amplitude of the corresponding anisotropy are very model-dependent, but a \%-level modulation is conceivable.
If the origin of cosmic rays is extragalactic, then their sources should be cosmologically distributed for all energies below the GZK attenuation, and no large-scale pattern should be detectable except perhaps for a CMB-like dipole with an amplitude of 0.6\%.

The AGASA collaboration found a $\sim 4\%$ amplitude modulation in right ascension (RA) in the  specific energy range $1 < E  <2$ EeV~\cite{agasa-ra}. The statistics accumulated by the Auger Observatory allows to study \%-level large-scale patterns, but this is challenging due to the difficulty of controlling the sky exposure of the detector with the corresponding accuracy. Indeed various acceptance effects, such as detector instabilities and weather modulations, can generate spurious modulations in the exposure.
We present here first results of complementary analyses in the EeV energy range, which set upper limits to a possible large-scale pattern. We also search for large anisotropies above 10 EeV, where it is widely believed that the CR sources are extragalactic.
All the analyses presented here use the high-level trigger (T5) events with a zenith angle $\theta \leq 60^{\circ}$, and the periods of unstable acquisition are removed.

\section{Fourier time analysis}

A modulation of the event rate over right ascension or sidereal time may be due to a true sky pattern, but also to the combined effect of diurnal and seasonal variations of the detector acceptance. To disentangle these effects, a widely used method is to compare sidereal modulations to solar and anti-sidereal ones~\cite{farley}. In the following analysis, we use the events recorded at the Auger Observatory from January 2004 to March 2007. For each frequency we compute the associated Fourier amplitude using the distribution of the times $t_i$ of the events. Rather than using directly $t_i$ we apply the analysis to the modified times $t_{\rm mod}(i) = t_i + {\rm RA}(i) - {\rm RA}_0(i)$, where $RA(i)$ is the right ascension of an event and $RA_0(i)$ is the associated local sidereal time~\cite{fourier_billoir}. Using such a definition, the sidereal modulation of $t_{\rm mod}$ is equal to the RA modulation of the event rate.

\begin{figure}
\begin{center}
\includegraphics[width=0.45\textwidth]{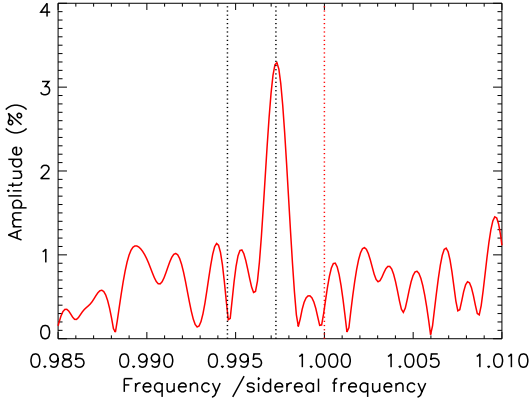}
\end{center}
\caption{Fourier analysis of the modified times $t_{\rm mod}$ for all events from 2004 to March 2007 ($E_{\rm median} \sim 0.6$ EeV). Frequencies are in unit of the sidereal one. The dotted lines represent, from left to right, the anti-sidereal, solar, and sidereal periods.}
\label{fig:fourier}
\end{figure}
 
The Fourier analysis on all the data ($5\times10^5$ events with a median energy $6\times 10^{17}$ eV), presented in Fig.~\ref{fig:fourier}, does not reveal a specific modulation at the sidereal frequency. A $3.2\%$ solar modulation of $t_{\rm mod}$ is observed, and can be interpreted as due to weather effects~\cite{weather_icrc} and to the array deployment and maintenance which take place during the day. Defining $T_{\rm exp}$ as the duration of the experiment, the frequency resolution is 
$\sim 1/T_{\rm exp}$, which is enough to resolve the solar and sidereal frequencies (as can be seen from the width of the solar peak in Fig.~\ref{fig:fourier}). 
The combination of the measured solar and yearly modulations of the acceptance may generate sidebands of equal amplitudes at the sidereal and anti-sidereal frequencies.

The analysis was repeated in 3 energy ranges with no significant sidereal modulation observed. In the energy range $1 < E < 3 $ EeV, with a statistics of almost $N=10^5$ events, a sidereal amplitude of 0.5\% is measured. The expected average noise from the Rayleigh distribution is $\sqrt{\pi/N} \simeq 0.6\%$, and an empirical estimate of this noise from the data at all frequencies (except for the solar band) gives a comparable value.

\begin{figure}
\begin{center}
\includegraphics[width=0.55\textwidth]{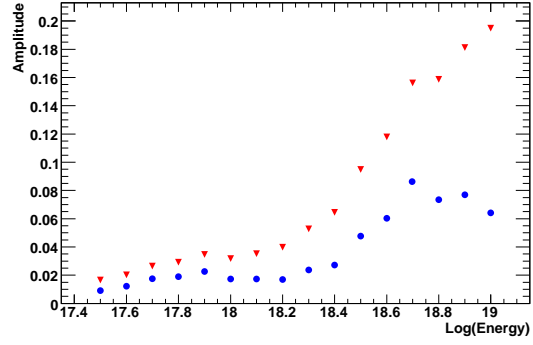}
\end{center}
\caption{Amplitudes (circles) and corresponding 95\% upper limits (triangles) of the first harmonic analysis in sidereal time vs. the energy threshold using the East-West method.}
\label{fig:ew}
\end{figure}

\section{East-West analysis}
To remove direction-independent phenomena, such as atmospheric and not-constant acceptance
effects, an interesting method consists in exploiting the differences in the numbers of counts between the
East-ward and the West-ward directions at a given time~\cite{eastwest}. Due to the symmetry of the array, the instantaneous East-ward and West-ward acceptances are practically equal. Assuming a harmonic modulation, the numbers of counts $E(t)$ and $W(t)$, from the East and West sectors respectively, with an average hour angle difference $\delta t$ with respect to the vertical, are related to the physical CR intensity $I(t)$ by $\frac{dI}{dt} \simeq \frac{E(t)-W(t)}{\sin \delta t}$. This is independent of the apparatus instabilities to the first order. The first harmonic analysis of $E(t)-W(t)$ gives an amplitude and a phase $(r_D,\phi_D)$, from which the amplitude and phase $(R_I,\phi_I)$ of the first harmonic of $I$ can be derived. 
The background distribution of $R_I$ follows the Rayleigh distribution, given for $N$ events by

$$ P(>R_I) = \exp\left[-\frac{N (R_I\, \sin \delta t)^2}{4}\right]$$

Using all the data from January 2004 to March 2007 with a declination $\delta \ge -70^{\circ}$, the solar day modulation of $4.2\%$ is reduced as expected, with the East-West procedure, to a $(0.9\pm 0.4) \%$ modulation. The raw sidereal amplitude is 0.6\%, and after the East-West procedure it is $(0.7\pm 0.4)\%$, corresponding to a chance probability of 24\%. From this measurement we derive a 95\% upper limit of $1.4\%$ on the amplitude of the first harmonic sidereal modulation at the median primary energy $6\times 10^{17}$ eV.
The results of the first harmonic analysis in sidereal time with the East-West procedure, as a function of energy threshold, is shown in Fig.~\ref{fig:ew}. All values are compatible with fluctuations, the minimum Rayleigh probability being 0.6\%.

\begin{table*}
\begin{center}
\begin{tabular}{|c|c|c|c|c|c|c|}
\hline
Energy & Declination & Number of events & Amplitude & Antisidereal  & 95\% Upper  \\ 
 $[$EeV$]$ & range & (expected noise) & (phase) & amplitude  & limit \\ \hline \hline
$1 < E < 3$ & All & 69641 (0.7\%) & 0.7\% ($330^{\circ}$) &  0.6\%  & 1.4\% \\ \hline
$1<E<3$ & $[-70^{\circ},-25^{\circ}]$ & 33470 (1.0\%) & 0.6\% ($350^{\circ}$) & 1.8\%  & 1.7\%  \\ \hline
$1<E<3$ & $[-25^{\circ},25^{\circ}]$ & 31144 (1.0\%) & 0.9\% ($310^{\circ}$) & 0.1\%  & 2.0\%  \\ \hline
$3<E<10$ & All & 7722 (2.0\%) & 1.2\% ($100^{\circ}$) & 1.1\% & 3.2\% \\ \hline \hline
$E>10$ & All & 1437 (4.7\%) & 3.5\% ($70^{\circ}$) & 4.2\% & 8.6\% \\ \hline
\end{tabular}
\label{table2}
\caption{Rayleigh analysis in different energy and declination ranges, for the periods 2005-2006 ($1<E<10$ EeV) and 2004-March 2007 ($E>10$ EeV). For each subset, we give the number $N$ of events and the Rayleigh noise $\sqrt{\pi/N}$; the Rayleigh amplitude and phase; the anti-sidereal amplitude derived from Fourier analysis on the modified time $t_{\rm mod}$; and the 95\% upper limit on a first harmonic modulation.}
\end{center}
\end{table*}

\section{Exposure-based study of the RA distributions}

The bidimensionnal sky exposure for a given set of events may be obtained through various ways~\cite{coverage}. In order to account for acceptance effects without absorbing real anisotropies, these methods use the observed zenith angle distribution of the events, but not the azimuth distribution. A small $60^{\circ}$-periodic azimuthal modulation of the low-energy event rate at zenith angles $\geq 45^{\circ}$, due to the array geometry, may however be taken into account. Different models for the time evolution of the acceptance were implemented. They can use the observed event rates, the array growth and dead times (using the instantaneous low-level trigger activity of individual tanks), as well as simple parameterisations of the pressure and temperature dependance of the acceptance~\cite{weather_icrc}. The derived exposure maps reproduce well the few \% RA modulation of the event rate which is observed when using a few months of data only. This modulation vanishes almost completely using the full 2-year data set of 2005-2006.
A systematic uncertainty on the predicted RA modulation of $\sim 0.4\%$ is derived from the differences between exposure estimates, for the 2-year period considered.

\begin{figure}[!t]
\resizebox{\hsize}{!}{\centering{\includegraphics[width=0.6\textwidth]{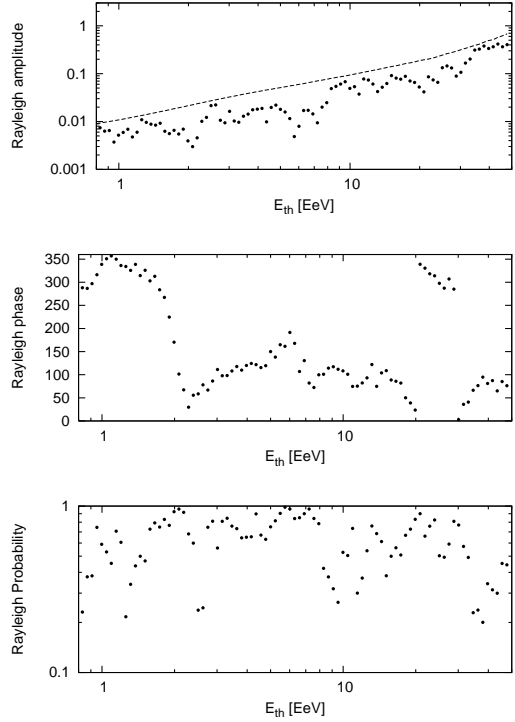}}}
\caption{Rayleigh scan over energy, correcting for the detector's geometrical acceptance. The dotted line represents the raw Rayleigh 95\% probability corresponding to the statistics associated to each energy threshold.}
\label{fig:ray}
\end{figure}

We describe the results of a Rayleigh analysis made using the 2005-2006 T5 data set with $\theta \le 60^{\circ}$. The exposure estimate models the time evolution of the acceptance using the detector growth and dead time data.
Fig.~\ref{fig:ray} presents a scan over energy of the Rayleigh amplitude, phase and significance. No significant modulation is detected throughout the scan.

The first lines of table 2 summarize the observed modulations in two energy ranges, $1<E<3$ EeV and $3<E<10$ EeV, as well as in two declination bands for the former energy range. In order to be conservative, upper limits are derived taking into account the observed residual RA modulation in the data. Given a number N of events, with an exposure-corrected RA modulation $A_{\rm data}$, one estimates from simulation, for each possible amplitude $A_0$, the distribution $p_{A0}(A)$ of the fitted amplitude with the available statistics N. The amplitude $A_0$ such that  $\int_{A_{\rm data}}^{\infty} dA\, p_{A0}(A) = 0.95$ is a relevant 95 \% upper limit to a RA modulation.

The Rayleigh amplitude is below 1\% at 1 EeV and an upper limit of 1.4\% is derived for $1<E<3$ EeV (there is a systematics of $0.4\%$ due to the exposure estimation, as mentionned above).

Above 10 EeV, the statistics accumulated so far is $\sim 10^3$ events, so that \%-level systematics are not an issue. Furthermore, as the acceptance is saturated, the use of the CIC method for energy calibration guarantees the flatness of the $\cos^2 \theta$ distribution of the events~\cite{spectrum_paper}. Therefore one can use an analytic exposure estimation, given for example in~\cite{Sommers:2000us}, and use all the T5 events with a reconstructed energy above $10^{19}$ eV recorded since 2004. The results of a Rayleigh analysis are displayed on the last line of table 2 :  an upper limit of 8.6\% on a RA modulation can be set.

\section{Conclusions}

Using the first years of Auger South data, we have searched for large-scale patterns in the sky distribution of cosmic rays, with special emphasis on those above 1 EeV. A series of complementary analyses has been used to study the systematics induced by detector acceptance variations and instabilities.
At EeV energies, the RA distribution is remarkably compatible with an isotropic sky, and an upper limit on the first harmonic modulation of 1.4\% in the energy range $1<E<3$ EeV is set. This does not confirm the 4\% RA modulation found by the AGASA experiment (however the sky regions covered by both experiments are not the same).
The results presented here concern the search for right ascension modulations, but studies are also going on to search for bidimensionnal patterns such as a possible dipole.

\bibliography{icrc0076}
\bibliographystyle{unsrt}

\end{document}